\documentclass[12pt]{article}

\usepackage{dcolumn}
\usepackage{bm}
\usepackage[latin1]{inputenc}
\usepackage[spanish,english]{babel}
\usepackage{amsfonts}
\usepackage{amssymb}
\usepackage{graphicx}
 \usepackage{setspace}
 \usepackage{verbatim} 
 \usepackage[normalem]{ulem}
 
\usepackage{color}
\usepackage{soul}

\usepackage[margin=3cm]{geometry}

\newcommand{\be}{\begin{equation}}
\newcommand{\ee}{\end{equation}}
\newcommand{\bea}{\begin{eqnarray}}
\newcommand{\eea}{\end{eqnarray}}

\begin{document}

\begin{titlepage}

\begin{center}
{\large \bf Absorption spectroscopy of quantum black holes with gravitational waves } 
\end{center}

\begin{center}

Ivan Agullo\\
{\footnotesize \noindent {\it {Department of Physics and Astronomy, Louisiana State University, Baton Rouge, LA 70803-4001, USA}}   }\\
Vitor Cardoso\\
{\footnotesize \noindent {\it {CENTRA, Departamento de F\'{\i}sica, Instituto Superior T\'ecnico -- IST, Universidade de Lisboa -- UL,
Avenida Rovisco Pais 1, 1049 Lisboa, Portugal}}   }\\
Adri\'an del Rio\\
{\footnotesize \noindent {\it {Institute for Gravitation and the Cosmos, Physics Department,
Penn State, University Park, PA 16802-6300, USA.}}   }\\
Michele Maggiore\\
{\footnotesize \noindent {\it {Departement de Physique Theorique and Center for Astroparticle Physics, Universite de Geneve, 24 quai Ansermet, CH-1211 Geneve 4, Switzerland}}   }\\
Jorge Pullin\\
{\footnotesize \noindent {\it { Department of Physics and Astronomy,Louisiana State University, Baton Rouge, LA 70803-4001, USA}}   }\\

\end{center}

\begin{abstract}

The observation of  electromagnetic radiation emitted or absorbed by matter was  instrumental in revealing the quantum properties of atoms and molecules in the early XX century, and constituted a turning-point in the development of  the quantum theory.
 Quantum mechanics changes dramatically the way radiation and matter  interact, making the probability of emission and absorption of light strongly frequency dependent, as clearly manifested in  atomic spectra. In this essay, we advocate that gravitational radiation can play, for the quantum aspects of black holes, a similar  role as electromagnetic radiation  did for  atoms,  and that  the  advent of gravitational-wave astronomy can bring this fascinating  possibility to the realm of observations. 

\end{abstract}

\vspace{0.8cm}

\begin{center}{\it Essay written for the Gravity Research Foundation 2021 Awards for Essays on Gravitation} \end{center}
\begin{center}Submitted on March 30, 2021 \end{center}
\end{titlepage}

That black holes (BHs) have similarities with atoms was  emphasized by Bekenstein five decades  ago \cite{Bekenstein:1974jk,Bekenstein:1997bt}. Not only their physical states are  characterized by a few numbers---mass $M$, spin $J$ and possibly electric charge--- but these parameters  may only take discrete values if one applies well established arguments by Bohr, Sommerfeld, Ehrenfest and others. Indeed, based on the observation that the area of BHs is an adiabatic invariant in general relativity (GR), Bekenstein concluded that the horizon area should be quantized,  and further argued that  the corresponding spectrum must be equally spaced in units of a fundamental quantum $\Delta A=\alpha\, \ell_{Pl}^2$, of the order of the Planck area $\ell_{Pl}^2=\frac{\hbar\, G}{c^3}\approx 10^{-70} \, {\rm m}^2$, where $\alpha$ is a constant of order one. Upon standard  quantization of angular momentum, one concludes that  the mass or energy spectrum must be discrete.  This quantization changes drastically the way BHs interact with classical radiation, discretizing the frequency of the waves they can absorb or emit \cite{Foit_2019},  in analogy to the atomic case. Bekenstein and Mukhanov \cite{Bekenstein:1995ju} explored potential  consequences  for the emission spectrum of BHs, i.e., for spontaneous Hawking radiation. The low Hawking temperature makes the observation of the predicted effects unreachable, at least for astrophysical BHs. However, the birth of GW astronomy offers an interesting alternative: to study the consequences of BH area quantization  for the {\em absorption spectrum}. This is the central topic of this essay.

An obvious question is the following: why should a discretization at the Planck scale of an astrophysical BH horizon  affect the low-frequency GWs that we observe in our detectors? The frequencies $ \omega_{\rm abs}$---or energies---that BHs can absorb are determined by  their  {\em mass} spectrum. In GR, the transition between two close states of definite mass  is  determined by the first law of BH mechanics \cite{Bardeen:1973gs}: 
\be
 \omega_{\rm abs}\equiv \frac{c^2\, \Delta M}{\hbar}= \frac{\kappa c^2}{8\pi {G} } \frac{\Delta A}{\hbar} +{\Omega_H}  \frac{\Delta J}{\hbar}\, , \label{BHtransition}
\ee
where $\kappa$ denotes the surface gravity and  $\Omega_H$ the horizon angular velocity. Since both $\kappa$ and $\Omega_H$ scale as $M^{-1}$, Eq.\ (\ref{BHtransition}) reveals that $\Delta M\propto  \frac{1}{M}$: the energy levels of a BH are not equally spaced when $A$ and $J$ are both uniformly discretized. 
Hence, energy levels get closer together for heavier BHs.  For instance,  the basic absorption frequency of  a non-rotating BH is
$ \omega_{\rm abs}=\frac{c^3}{32\pi G\, M_{\odot}}\, \alpha \, \frac{M_{\odot}}{M}$,
where we have expressed $M$ in units of the solar mass $M_{\odot}$. The interesting observation is that the combination of constants $\frac{c^3}{32\pi G\, M_{\odot}}$ is approximately $2$ kHz---this is not  a numerical accident: parametrically this is the same as the frequencies of the quasi-normal modes (QNMs) discussed below.
Thus, the large mass of typical astrophysical BHs  is responsible for translating Planck-scale discretization of the BH area to frequencies within the window of GW detectors.

Recent investigations have revealed two observable channels in the coalescence of binary black hole systems that  can inform us on the discrete energy spectrum of BHs, related to the ringdown and the inspiral phases \cite{Maselli:2017cmm, Cardoso:2019apo, Agullo:2020hxe}.  

The ringdown phase is accurately described by a  perturbed Kerr BH, which dissipates its perturbations via its characteristic QNMs. 
 Most of the energy carried by the QNMs moves outwards to GW detectors, but a non-negligible fraction propagates inwards, towards the horizon. This is a remarkably monochromatic  radiation, because it is dominated by the quadrupolar mode $\ell=2$, $m=2$ and the fundamental tone $n=0$.  The oscillation frequency  of this wave is  $M{\rm Re}\, \omega_{022}\simeq 1.5251-1.1568\, (1-a)^{0.1292}$,  where $a=J/M^2$.  If this frequency does not match any transition of the BH mass spectrum, Eq.~(\ref{BHtransition}), the probability of absorption will be suppressed. The incoming radiation will then start propagating outwards. Once it interacts with the light ring, a portion will be scattered back, and the process will repeat in time.  As a result, an external observer will see an initial GW burst followed by a set of echoes with increasingly smaller amplitudes \cite{Cardoso:2016rao, Cardoso:2019apo, Cardoso:2019rvt}. The values of these amplitudes depend on the exact absorption properties of the BH---determined by unknown microscopic physics---and on the transmissivity of the potential barrier. Reasonable estimates indicate that the amplitude of the first echo could be as large as a percent of the initial GW front.  The ability to detect these echoes depends on our capacity to produce faithful templates.  For  existing phenomenological families of echo waveforms, Bayesian analysis in the LIGO/Virgo data do not find evidence for echoes with amplitude $0.1-0.2$ times to the original signal  peak~\cite{Westerweck:2017hus}. Observational  constraints will improve significantly in the near future with  3G detectors such as the Einstein Telescope (ET)~\cite{Maggiore:2019uih} and the   space mission LISA~\cite{Audley:2017drz}.

The second possibility to test the area quantization hypothesis relies on the inspiral. During this stage  the system emits GW radiation, again dominated by the quadrupolar mode, with frequency given  by the binary orbital angular velocity $\Omega$,  $\omega\approx 2\Omega $. Classically, the individual BHs absorb a portion of these waves, which induce tidal forces that distort  the  horizons.  But as the distorted BHs rotate, energy is dissipated gravitationally, and transferred to the inspiral dynamics. This phenomenon is known as tidal heating \cite{Cardoso:2019apo}. Now, because the orbital angular velocity is considerably low during inspiral, the frequency of these GWs is smaller than the lowest absorption frequency of the individual BHs, $\omega_{\rm abs}\sim 2\Omega_H $  (see Eq.~(\ref{BHtransition}) and comments below).  Consequently, the GW absorption is expected to be suppressed---this is analog to the familiar frequency threshold in the photoelectric effect---modifying  the orbital evolution of the binary as compared to the classical prediction. One can study this effect by introducing  an absorption parameter   $\gamma$ in the  waveform (multiplying the 2.5 PN$\times \log v$ GR term, \cite{Maselli:2017cmm}). Quantization of area  decreases $\gamma$ relative to its value for classical BHs, $\gamma_{\rm class}=1$.  The analysis of ~\cite{Maselli:2017cmm} reveals that advanced detectors such as LISA and the ET will reach the desired sensitivity to discriminate among these values of $\gamma$.

Although these  effects are  consequences of the quantization of the BH  energies, this discreteness alone is not sufficient. The energy levels are  determined  from the area and angular momentum quantum numbers $n$ and $j$ by $M_{n,j}=\sqrt{\hbar}\, \sqrt{ \frac{\alpha\, n}{16 \pi}+\frac{4\pi j^2}{\alpha\, n}}$. This spectrum consists not only of Schwarzchild states, but it also includes all spinning configurations.  It is a highly irregular and crowded spectrum which, under consideration of the linewidths (see below), approaches a continuum. Nevertheless, as recently pointed out in \cite{Agullo:2020hxe}, conservation of angular momentum introduces constraints, or ``selection rules'',  which, as in atomic physics, restrict the energy levels  that the BH can transition to when it interacts with a GW  mode $(\omega, \ell, m)$.  As previously emphasized, the GWs generated  during the inspiral and  ringdown stages are dominated by   quadrupolar modes ($\ell=2$, $m=2$), that  single out BH energy levels  differing by $\Delta j=2$. These levels form a small and simple subset of the spectrum $M_{n,j}$. Furthermore, since these GWs are also highly monochromatic, unless their frequency matches  one of the absorption frequencies of the BH,  the probability of absorption would be suppressed, giving rise to the effects described above. Therefore, it is the combination of energy quantization and angular momentum conservation  that gives rise to  observable effects.

Another important point to take into account is the width $\Gamma$  of the energy levels: 
 observable effects exist only if $\Gamma$ does not make the relevant energy levels discussed above to overlap \cite{Coates:2019bun,Agullo:2020hxe}.  This linewidth can be estimated as $\Gamma=\hbar/\tau$, where $\tau$ is the spontaneous decay rate, i.e., the  timescale associated with  Hawking emission.  This can be computed following Page's  calculations~\cite{Page:1976ki}. Reference \cite{Agullo:2020hxe} has  computed the ratio $R(a, \alpha)=\Gamma/(\hbar \,  w_{\rm abs}|_{\Delta j=2})$, and has showed it is a function of the BH spin $a$ and the size $\alpha$ of the fundamental area gap; the explicit dependence on  BH mass $M$ cancels out. $R(a, \alpha)$ can be large if $\alpha$ is sufficiently small---in fact $R$ diverges in the limit $\alpha\to0$, in which the BH recovers the continuous energy spectrum. The interesting quantity is therefore $\alpha_{\rm crit}(a)$, the value of $\alpha$ below which the relevant energy levels in a binary coalescence overlap, $R(a,\alpha_{\rm crit}(a))=1$. This  quantity is
\be \alpha_{\rm crit}(a)=0.0842+0.2605\, a^2+0.0320\, e^{5.3422\, a^3}\, ,\ee
accurate to within $2\%$ for $a<0.9$. As an example, for $a \approx 0.7$---the spin of the remnant BH found in a large fraction of the observed mergers---one obtains $\alpha_{\rm crit}=0.42$. This is one order of magnitude below $4 \log{2}$, the smallest value of $\alpha$ that has been discussed in the literature. Hence, for reasonable values of  $\alpha$ and BH spins, the relevant energy levels do not overlap.

\begin{figure}[h!]
\centering
\includegraphics[scale=0.53,trim=0cm 0cm 0cm 0cm]{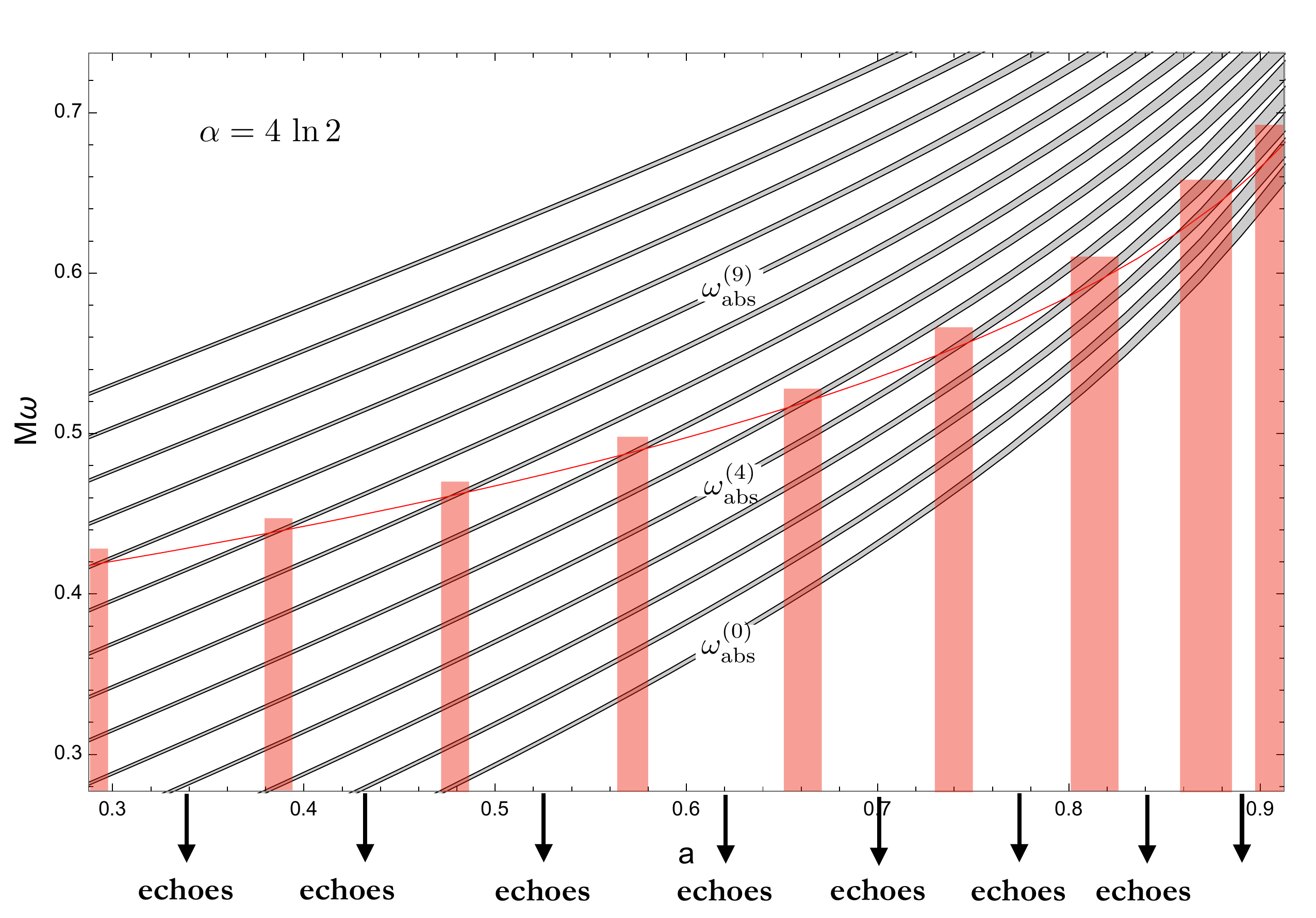}
\caption{Black lines: absorption frequencies associated with the relevant energy levels of BHs in a binary merger. The (gray) thickness of the black lines measures the width $\Gamma$ of the spectral lines. The red line is the oscillation frequency of the dominant QNM, as a function of the BH spin $a$. This plot does not change with the mass $M$ of the final black hole, since all quantities plotted scale in the same manner. The intersection of the red and black lines, highlighted with vertical red bands, correspond to values of $a$ for which the probability of absorption of the dominant QNM is close to one. Echoes are therefore expected in the regions in between intersections. This plot is obtained using the smaller value of $\alpha$ that has been discussed in the literature. Larger values of  $\alpha$ increase the range of $a$ for which echoes are expected.}
\label{fig}
\end{figure}

A summary of this discussion is represented in Fig. \ref{fig} (see \cite{Agullo:2020hxe} for more details). The spin of BH's enriches the phenomenology in an unforeseen manner.  In particular, the analysis shows that an observation of echoes for binary mergers sampling a large enough range of the BH spin $a$,  provides a way to test the BH area spectrum and to determine the fundamental constant $\alpha$. 

As surprising as it may sound,  that GW radiation can inform us about quantum aspects of BHs is a plausible possibility, which certainly deserves further scrutiny. We encourage  the quantum gravity community to derive concrete predictions, including GWs templates, which could  be contrasted with current and future GW observations.\\

\noindent{\bf{\em Acknowledgments.}}
%
V. C. acknowledges financial support provided under the European Union's H2020 ERC 
Consolidator Grant ``Matter and strong-field gravity: New frontiers in Einstein's 
theory'' grant agreement no. MaGRaTh--646597. 
This project has received funding from the European Union's Horizon 2020 research and innovation 
programme under the Marie Sklodowska-Curie grant agreement No 690904.
We thank FCT for financial support through Project~No.~UIDB/00099/2020 and through grant PTDC/MAT-APL/30043/2017.
The authors acknowledge networking support by the GWverse COST Action 
CA16104, ``Black holes, gravitational waves and fundamental physics.'' 
AdR. acknowledges support under NSF grant PHY-1806356 and the Eberly Chair funds of Penn State. 
MM is supported by the  Swiss National Science Foundation and  by the SwissMap National Center for Competence in Research. 
IA  is supported by the NSF CAREER grant PHY-1552603 and by the Hearne Institute for Theoretical Physics. 
JP is supported by grant NSF-1903799, by the Hearne Institute for Theoretical Physics and CCT-LSU.

\bibliographystyle{utphys}
\bibliography{References}

\end{document}